\documentclass[a4paper,11pt]{article}
\usepackage{jcappub} 

\usepackage{bm}

\usepackage{array}

\title{\boldmath Measuring the growth index in the $w_0w_a$CDM background}

\title{Measuring the growth index in the $w_0w_a$CDM background}
\author[a]{Í. B. S. Cortês }
\emailAdd{icaro.cortes.710@ufrn.edu.br}
\author[b]{Léo G. Medeiros}
\author[b]{Ronaldo C. Batista}
\affiliation[a]{Departamento de Engenharia de Computação e Automação, Centro de Tecnologia,
Universidade Federal do Rio Grande do Norte, Natal, Rio Grande do
Norte, Brazil.}
\affiliation[b]{Escola de Ciências e Tecnologia, Universidade Federal do Rio Grande
do Norte, Natal, Rio Grande do Norte, Brazil.}

\abstract{
We produce a measurement of the growth index, $\gamma$, which is used to parametrize the scale-independent growth of linear matter perturbations on sub-horizon scales, for the $w_0w_a$CDM background model. In 2023, assuming the $\Lambda$CDM background, the value $\gamma =0.633^{+0.025}_{-0.024}$ was obtained, excluding the General Relativity value of $\gamma=0.55$ at $3.7\sigma$ statistical significance. In 2025, an analysis of baryon acoustic oscillation data from the Dark Energy Spectroscopy Instrument (DESI) demonstrated a preference for a dynamical dark energy over the cosmological constant model, $\Lambda$, ranging from $2.8-4.2\sigma$, depending on the external datasets used. Therefore it is important to determine how the $\gamma$ measurement is modified in the context of dynamical dark energy and verify whether the tension with the expected theoretical value persists. Considering data from Planck cosmic microwave background anisotropies, $f\sigma_{8}$, baryon acoustic oscillation from DESI and supernova Ia luminosity distances from Phanteon+, we find $\gamma_{w_0w_a}=0.633\pm0.024$, essentially the same value found for the $\Lambda$CDM background. This result excludes the General Relativity value for the $w_0w_a$CDM background evolution at $3.4\sigma$ level. We also show that the $\gamma$ posteriors across many datasets combinations are consistent with each other and that main driver of the "$\gamma$ tension" is the $f\sigma_{8}$ dataset.
}

\begin{document}
\maketitle
\flushbottom

\section{Introduction}
The growth index parametrization is a particularly simple and powerful tool to study the growth linear matter perturbations on sub-horizon scales and at late times. The parametrization expresses the growth rate of matter perturbation, $f(a)$, as \cite{Linder:2005in,Amendola2013,Huterer2018}
\begin{equation}
f(a) \equiv \frac{d\ln \delta_{m}(a)}{d\ln a}\simeq\Omega^{\gamma}_{m}(a),
\label{gamma-defition}
\end{equation}
where $\gamma$ is the growth index, $a$ the scale factor, $\Omega_{m}(a)=\bar{\rho}_m(a)/\bar{\rho}_c(a)$ the matter density parameter, $\bar{\rho}_c(a)$ the critical energy density and $\delta_m(a)=\delta\rho _m (a)/\bar{\rho}_m(a)$ the linear scale-independent matter density contrast. 

It is well known that, in many cosmological models, the growth rate is scale-independent and can be accurately described by a constant $\gamma$. For the $\Lambda$ cold dark matter ($\Lambda$CDM) model, we have $\gamma\simeq0.55$ \citep{Wang:1998gt,Linder:2005in}, which is usually referred as the General Relativity value. More precisely, the $\Lambda$CDM model constrained only with Cosmic Chronometers data predicts $\gamma=0.5514\pm0.0010$, with parametrization errors about $0.2\%$ across the large parameter space allowed by this dataset \citep{Cortes2025}. For dynamical smooth dark energy  (DE) model described by the $w_0w_a$ parametrization \cite{Chevallier:2000qy,Linder:2003dr}, in which the equation of state parameter (EoS) is given by

\begin{equation}
    w(a)=w_0+w_a(1-a),
\end{equation}
$\gamma$ has a weak dependence on the equation of state \citep{Linder:2005in,Cortes2025}. But for more complex EoS evolution or in the case of clustering DE, $\gamma$ can present stronger variations, \citep{Batista2014a,Cortes2025}.

In 2023, assuming the $\Lambda$CDM background and making use of Planck 2018 data release \citep{Aghanim:2018oex} (PL18) cosmic microwave background (CMB) anisotropies, $f\sigma_8$, barion acoustic oscillations (BAO), Dark Energy Survey Year 1 galaxy clustering datasets, ref. \cite{Nguyen:2023fip} measured $\gamma=0.633^{+0.025}_{-0.024}$, excluding the theoretical value at $3.7\sigma$ level. Later, the Atacama Cosmology Telescope collaboration found $\gamma=0.630\pm 0.023$, when making use of their Data Release 6 CMB spectra, Dark Energy Spectroscopic Instrument (DESI) Year-1 BAO and the same $f\sigma_8$ dataset used in the 2023 measurement \citep{AtacamaCosmologyTelescope:2025nti}. This discrepancy between the measured $\gamma$ and the theoretical value predicted by the $\Lambda$CDM model can be called as $\gamma$ tension, which can be associated with the well-known $S_8$-tension, where $S_8=\sigma_8\sqrt{\Omega_m/0.3 }$ \cite{Nguyen:2023fip,Abdalla:2022yfr,Cai:2026swf}, where $\sigma_8$ is the amplitude of matter perturbations now at $8\mbox{Mpc}h^{-1}$,  $\Omega_m$ is the matter density parameter today and $h$ is related to the Hubble constant as $H_0=100h\mbox{km/s/Mpc}$. 

A couple of years after this $\gamma $ determination, the analysis of DESI BAO data release 2 (DR2), in conjunction with CMB and super novae (SNe) Ia datasets, indicated a preference for dynamical DE over $\Lambda$ ranging from $2.8$ to $4.2\sigma$, depending on the SNe Ia dataset used \citep{DESICollaboration2025}. A natural question in this situation is whether the evidence for a non-standard growth index is affected when the late background evolution is dictated by dynamical DE. 

In ref. \citep{Cortes2025}, it was shown that smooth DE models described by the $w_0w_a$ parametrization have strong difficulties in providing $\gamma> 0.6$ and a preliminary analysis indicated that measured $\gamma$ for this background should be very similar to one in the $\Lambda$CDM case. Moreover, the same study indicated that clustering DE \citep{Batista:2021uhb}, models could provide slightly larger $\gamma$, specially when the EoS parameters are constrained by DESI DR2 BAO, which prefers a phantom behavior at intermediate redshifts. However, a robust comparison between the theoretical $\gamma$ values for dynamical DE models and the actually measured $\gamma$ could not be done because they assume distinct backgrounds evolutions. 

Beyond the usual DE scenario, but still assuming the $\Lambda$CDM background, the high $\gamma$ value can be interpreted as a time-variation of the Newton's gravitational constant with a tight linear correlation between $\gamma$ and the parameter describing the time variation of $G$ \cite{Cortes:2025wni}. Such behavior is not usual in modified gravity models, but could be explained by recently proposed model, ref. \citep{Pookkillath:2026wyg}.

In this work, we make use of several combinations of PL18, $f\sigma_8$, DESI DR2 BAO and Pantheon+ SNe Ia datasets to constraint the value of $\gamma$ assuming that the DE EoS is described by the $w_0w_a$ parametrization. We find that, according to earlier expectations, the measured value of $\gamma$ for the $w_0w_a$CDM background is quite similar to the $\Lambda$CDM one found in refs. \cite{Nguyen:2023fip, AtacamaCosmologyTelescope:2025nti}. Therefore, even in the case of dynamical DE preferred by DESI BAO DR2, the $\gamma$ tension persists. We also analyze whether the $\gamma$ determinations based on several datasets are consistent, finding that for both background evolutions all data combinations considered provide compatible results. 

This paper is organized as follows. In the next section we present the methods and datasets used. In section \ref{sec:results} we present the results of our analysis with several datasets combinations. In section \ref{sec:conclusion}, we conclude.

\section{Models, codes and datasets\label{sec:method}}
In this work, we analyze two background cosmological evolutions, $\Lambda$CDM and $w_0w_a$CDM, with linear matter growth function characterized by a free constant $\gamma$. For the $\Lambda$CDM case, besides consistency checks with results from refs.  \citep{Nguyen:2023fip,AtacamaCosmologyTelescope:2025nti}, we want to understand whether the determination of $\gamma$ is affected by the inclusion of DESI DR2 BAO. Our main goal is to measure the $\gamma$ value in $w_0w_a$CDM background, which is more flexible and capable to better describe DESI data with respect to $\Lambda$CDM. We consider a universe with flat spatial section, thus the Friedmann equation is given by 
\begin{equation}
    H^2(a)=H^2_0\left(\Omega_{\gamma} a^{-4}+\Omega_{\nu}(a)+\Omega_{cb} a^{-3} +\Omega_{de}(a)\right)\,
\end{equation}
where $\Omega_{\gamma}$ is the photon energy density parameter today, $\Omega_{\nu}(a)$ is the time-dependent density parameter for massive neutrinos (we assume $\sum m_{\nu}=0.06\mbox{eV}$), $\Omega_{cb}=\Omega_{c}+\Omega_{b}$ is the sum of CDM and baryon density parameters today and $\Omega_{de}(a)= \Omega_{de} a^{-3(1+w_0+w_a)}\exp(-3w_a(1-a))$ is the time-dependent DE energy density parameter. Since we assume flat spatial section, the sum of all energy parameters must equals unity, thus we have $\Omega_{de}=1-\Omega_{\gamma}-\Omega_{\nu}-\Omega_{cb}$. 

Given that the transition for relativistic to non-relativistic behavior of massive neutrinos with $\sum m_{\nu}=0.06\mbox{eV}$ occur near $z=100$ \citep{ParticleDataGroup:2022pth}, for the computation of the parametrized growth rate, which is relevant at much lower $z$, in eq. (\ref{gamma-defition}), we define the time-dependent matter density parameter as
\begin{equation}
    \Omega_m(a)=\Omega_{cb} a^{-3}+\Omega_{\nu}(a)\,.
\end{equation}

In the case of dynamical DE, care must be taken with DE perturbations. We want to leave any data preference for non-standard perturbative behavior to be absorbed by $\gamma$, avoiding modifications on it induced by DE perturbations, while allowing for phantom crossing behaviors of the EoS. In order to do so, we assume that DE is described by the parametrized post-Friedman framework (PPF) \citep{Hu2007,Fang:2008sn}, which assumes that DE perturbations are much smaller than matter ones. In the fluid interpretation, this is nearly equivalent to assume unity sound speed for DE perturbations. More details on this point will be given in the next subsection.

\subsection{Codes}

To compute the cosmological observables of interest, we use \texttt{CAMB\_GammaPrime\_Growth} \citep{Nguyen:2023fip} (a modified version of \texttt{CAMB} \citep{Lewis2000,Howlett2012} that implements the modified growth of matter perturbations parametrized by $\gamma$), \texttt{Cobaya} as Markov Chain Monte Carlo sampler \citep{Torrado2020, Lewis2002} and \texttt{GetDist} \citep{Lewis2019} to analyze the posteriors and produce the plots. We have made a small modification in the \texttt{CAMB\_GammaPrime\_Growth} code to allow the sampling of $\gamma$ when using the $w_0w_a$ parametrization. We indirectly make use of the code \texttt{GIMBAL} \footnote{Available at \href{github.com/icarob-eng/gimbal\_pub}{github.com/icarob-eng/gimbal\_pub}.}, presented in ref. \citep{Cortes2025} when referring to the theoretical $\gamma$ predictions that will be compared with the measurements. We set \texttt{CAMB\_GammaPrime\_Growth} to make use of \texttt{HMCode2020} \citep{Mead:2020vgs} to compute the nonlinear matter power spectrum necessary to calculate lensing effects on the primary CMB anisotropies. We note that, as discussed in \citep{Nguyen:2023fip}, in this setup, \texttt{HMCode2020} incorporates the impact of $\gamma$ only via the modification on the linear matter power spectrum.

We performed the samplings assuming the priors for primary parameters listed in table \ref{tab:priors}. The reference distributions and sampling scales where the default values given by \texttt{Cobaya} input generation tool and some chains where initialized with the covariance matrix of the CMB-only chains, with their respective parametrizations. All samplings where performed with 6 parallel chains and a Gelman-Rubin statistic of $R-1<0.01$ as the convergence criteria, ending with a acceptance rate of $~35\%$ for each run. The chain lengths vary from $\sim10\,000$ for the simpler parametrizations up to $\sim180\,000$ for the more complex cases. To analyse the posteriors, we discarded the first $30\%$ samples as a "burn-in" in the \texttt{GetDist} code.

\begin{table}[h]
\centering
\begin{tabular}{cc}
\hline 
Parameter & Prior\tabularnewline
\hline
$\Omega_{b}h^{2}$ & $\mathcal{U}[0.005,0.1]$\tabularnewline
$\Omega_{c}h^{2}$ & $\mathcal{U}[0.001,0.99]$\tabularnewline
$100\theta_{\text{MC}}$ & $\mathcal{U}[0.5,10]$\tabularnewline
$n_{s}$ & $\mathcal{U}[0.8,1.2]$\tabularnewline
$\log(10^{10}A_{s})$ & $\mathcal{U}[1.16,3.91]$\tabularnewline
$\tau_{\text{reio}}$ & $\mathcal{U}[0.01,0.8]$\tabularnewline
\hline
$w_{0}$ & $\mathcal{U}[-3,1]$\tabularnewline
$w_{a}$ & $\mathcal{U}[-3,2]$\tabularnewline
$\gamma$ & $\mathcal{U}[0,2]$\tabularnewline
\hline
\end{tabular}
\caption{Uniform priors adopted in all cases, when relevant.}
\label{tab:priors}
\end{table}

The specific implementation of the \texttt{CAMB\_GammaPrime\_Growth} code computes the impact of $\gamma$ on the matter growth function, which is then used to rescale the linear matter power spectrum within \texttt{CAMB}’s \texttt{HaloFit} module. The resulting nonlinear matter power spectrum is used to calculate CMB lensing effects, while leaving the primary CMB anisotropies unchanged. 

To describe the DE models, we use the PPF module in \texttt{CAMB}, which allows for phantom crossing of the EoS and assumes that DE perturbations are much smaller than the matter ones. As is well-known, the unity sound speed makes DE perturbations irrelevant on small-scales, but might have some impact on larger scales, specially via the integrated Sachs-Wolf effect on CMB large scale anisotropies. 

To compute the linear evolution of matter perturbations, which will be used to determine the $f\sigma_8$ observable, the ref. \citep{Nguyen:2023fip}'s analysis uses the modified nonlinear power spectrum at $k=2\times 10^{-4}h/{\mbox Mpc}$, where nonlinear corrections are negligible. We checked that changing this scale to $k=10^{-3}h/{\mbox Mpc}$ has negligible impact on the parameters posteriors, thus showing that DE perturbations in models described by the $w_0w_a$ parametrization with the PPF description have virtually no impact on the $\gamma$ constraints.

Moreover, as a consistency check for the theoretical $\gamma$ values that we quote in this work, which where produced in ref. \citep{Cortes2025} with the \texttt{GIMBAL} code, we compare them with the values derived directly from $f\sigma _8$ functions produced by \texttt{CAMB\_GammaPrime\_Growth}. We found that the differences in the determination of $\gamma$ from \texttt{CAMB} to \texttt{GIMBAL} for smooth (using PPF) and clustering DE (with $w(a)>-1$ and using the fluid model) are typically less than $0.6\%$.

It is important to note that smooth DE model described by $w_0w_a$ parametrization and constrained by CMB priors and DESI present an intrinsic $\gamma$ variation around the $0.55$ value, ref. \citep{Cortes2025}. However, their $1\sigma$ limits are one order of magnitude lower than the current best constraints we will report in this work. Therefore, this theoretical dispersion associated with background variation beyond $\Lambda$CDM has negligible impact on the determination of $\gamma$ and corresponding tensions.  We recall, however, that DE model with null sound speed can present a more significant variation on the mean values of $\gamma$ and with more dispersion than the smooth counterparts \citep{Batista2014a,Cortes2025}.

\subsection{Datasets}
To measure $\gamma$ we make use of the following datasets and their respective short-names:
\begin{itemize}
    \item {\bf PL18}: the baseline CMB data from Planck's 2018 data release. This set is comprised of low-$\ell$ EE and TT likelihoods, the high-$\ell$ TTTEEE \texttt{plik} likelihood \citep{Aghanim:2019ame}, as well as the lensing likelihood \texttt{clik} \citep{Aghanim:2018oex}.

    \item {\bm{$f\sigma_8$}}: a set of the same $f(z)\sigma_8(z)$ data points used in ref. \citep{Nguyen:2023fip} with diagonal covariance matrix, where $f(z)$ is the growth rate of matter perturbations and $\sigma_8(z)=\sigma_8D(z)$ is the root-mean-square of the matter power spectrum smoothed on $8\mbox{Mpc}h^{-1}$ scale times the growth function, $D(z)=\delta_m(z)/\delta(z=0)$. These data points are compiled from several galaxy surveys and obtained either via peculiar velocities or redshift space distortions. 

    \item {\bf DESI}: BAO measurements incorporated in the likelihood of all tracers from Dark Energy Spectroscopy Instrument second data release \cite{DESI:2025zpo,DESI:2025zgx}.

    \item {\bf PP}: the Pantheon+ type Ia supernovae likelihood excluding \texttt{SH0ES} data \citep{Brout:2022vxf}. To save computational time and limit dataset combinations, we only consider this SNe Ia dataset, which is the one that produces the smaller tension with the $\Lambda$CDM hypothesis in the DESI DR2 analysis \citep{DESICollaboration2025}.

    \item {\bf CMB-P}: a triplet of correlated priors on $(\theta_*, \omega_b,\omega_{bc})$, where $\theta_*$ is the angular scale of the first CMB acoustic peak, $\omega_b=h^2\Omega_b$ and $\omega_{cb}=h^2\Omega_{cb}$, derived from CMB anisotropies spectra and constructed to marginalize over late-time effects such as integrated Sachs-Wolf effect and CMB lensing, as determined by ref. \citep{Lemos:2023xhs}.
\end{itemize}

\section{Results\label{sec:results}}
Our main results are listed in table \ref{tab:result}. As can be seen, the $\gamma$ constraints for the $\Lambda$CDM background are quite similar to those reported in Refs. \citep{Nguyen:2023fip,AtacamaCosmologyTelescope:2025nti}. In particular, when combining PL18, $f\sigma_8$, DESI and PP dataset we find 
\begin{equation}
\gamma_{\Lambda}=0.635\pm0.024.
\label{gamma-lambda}
\end{equation}

\begin{figure*}
\centering
\includegraphics[width=0.49\linewidth]{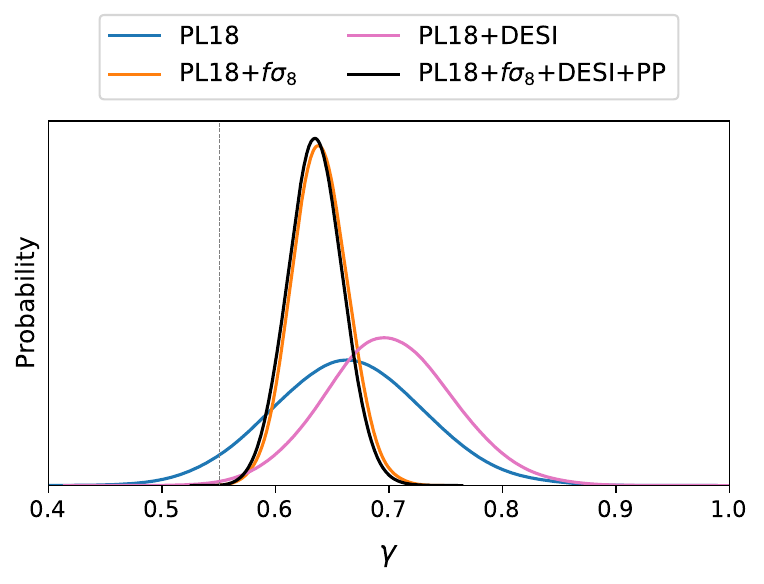}
\includegraphics[width=0.49\linewidth]{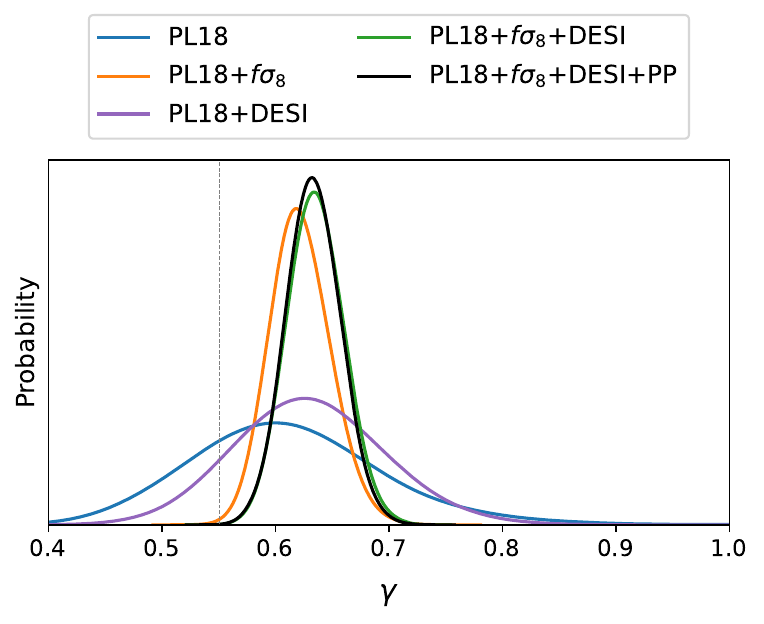}
\caption{Marginalized one-dimensional distribution of $\gamma$ in the $\Lambda$CDM background (left panel) and in the $w_0w_a$CDM one (right panel). Dashed lines indicates $\gamma=0.55$.}
\label{fig:gamma-dist}
\end{figure*}

\begin{figure*}
\centering
\includegraphics[scale=0.80]{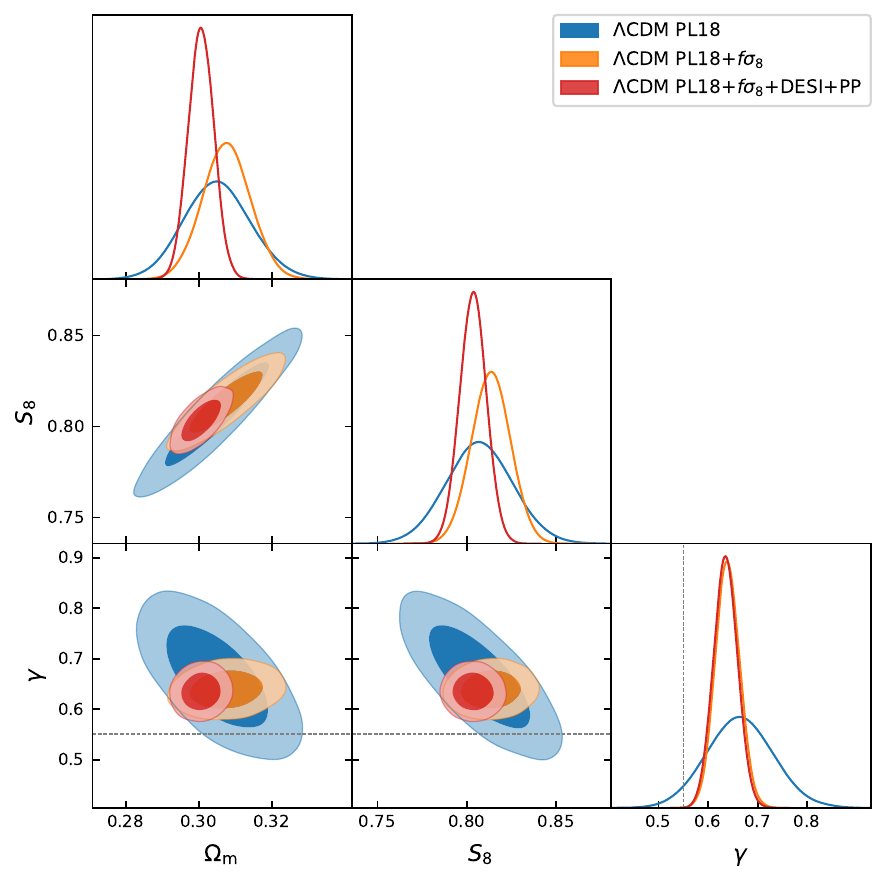}
\caption{Posterior distributions for $\gamma$ and selected parameters for three datasets combinations for the $\Lambda$CDM background. The contours enclose $68\%$ and $95\%$ of posteriors distributions. Dashed lines indicates $\gamma=0.55$.}
\label{fig:trig-LCDM}
\end{figure*}

In figure \ref{fig:gamma-dist}, we show the dependence of the $\gamma$ determination on the several datasets combinations. It is visually clear that all $\gamma$ posteriors distributions are compatible and that most of the information that constrains it is comes from the $f\sigma_8$ dataset. For the $\Lambda$ model, the inclusion of background data (DESI+PP) over the PL18+$f\sigma_8$ dataset has negligible impact on the determination of $\gamma$. In figure \ref{fig:trig-LCDM}, we also show some posteriors for $\gamma$, $\Omega_m$ and $S_8$.

As can be seen in table \ref{tab:result} and figure \ref{fig:gamma-dist}, in the case of $w_0w_a$CDM background, the background data has a more relevant impact on the $\gamma$ determination. The PL18 dataset provides poor constrains on the EoS parameters, which translates into a larger $\gamma$ uncertainty with respect to the case for $\Lambda$CDM background. The inclusion of $f\sigma_8$ data improves the $\gamma$ determination substantially. The inclusion of background data, which provides much better constraints on $w_0$ and $w_a$, now gives a slightly better $\gamma$ determination with respect to the latter combination. Our best constraint using PL18, $f\sigma_8$, DESI, and PP datasets is 
\begin{equation}
\gamma_{w_0w_a}=0.633\pm0.024\,.
\label{gamma-w0wa}
\end{equation}

As can be seen, the $\gamma$ measurements for $\Lambda$CDM and $w_0w_a$CDM background models are virtually the same. This result was qualitatively predicted in ref. \citep{Cortes2025}, when estimating how $\gamma$ should change when fitting the growth rate for the best-fit $\Lambda$CDM background with the $w_0w_a$CDM background constrained by CMB priors and DESI DR2 BAO. A possible interpretation for this result is the mirage DE model that states that the relation $w_a\simeq -3.66(1+w_0)$, which is essentially the correlation line favored by DESI data, mimics the $\Lambda$CDM background evolution \citep{Linder:2007ka,DESI:2025fii}.  

\begin{figure*}
\begin{centering}
\includegraphics[scale=0.60]{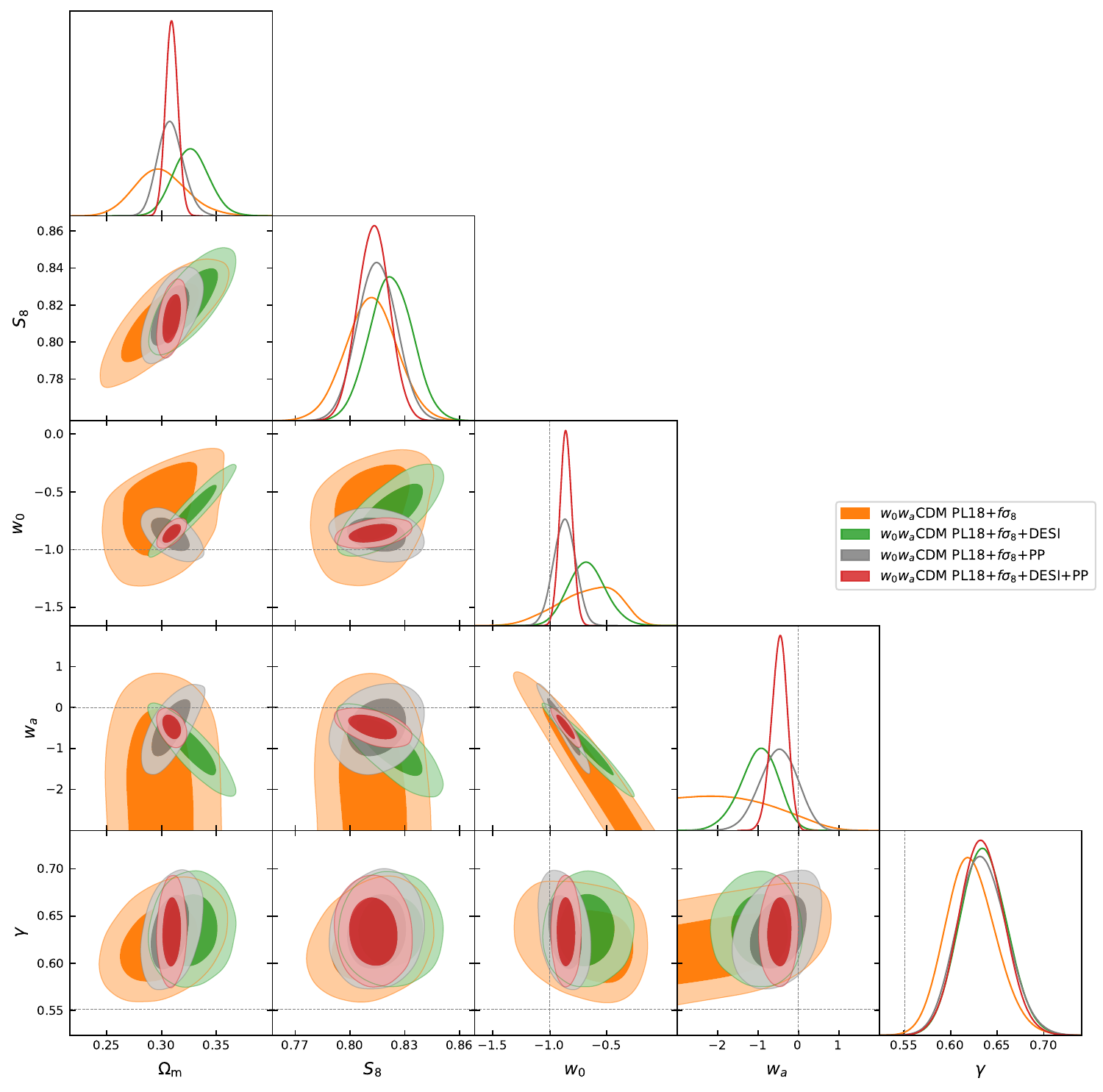}
\par\end{centering}
\caption{Posterior distributions for $\gamma$ and selected parameters for four datasets combinations for the $w_{0}w_{a}$CDM background. The contours enclose $68\%$ and $95\%$ of posteriors distributions. Dashed lines indicates $\gamma=0.55$, $w_0=-1$ or $w_a=0$.}
\label{fig:trig-w0wa}
\end{figure*}

In figure \ref{fig:trig-w0wa} we show the posteriors for $\gamma$, $w_0$, $w_a$, $\Omega_m$ and $S_8$ for some selected datasets. Interestingly, $\gamma$ has no strong correlations with the background parameters $w_0$, $w_a$ and $\Omega_m$ and the standard value of $\gamma\simeq0.55$ is beyond the $2\sigma$ regions of all joint distributions. This also supports the finding of nearly same $\gamma$ values for the two background models. From table \ref{tab:result}, we can see that the determinations of $w_0$ and $w_a$ with the PL18+$f\sigma_8$+DESI+PP are compatible to the results from ref. \citep{DESICollaboration2025} ($w_0=-0.838\pm0.055$ and $w_a=-0.62^{+0.22} _{-0.19}$), with a very small improvement on their precision. However, we note that the inclusion of $f\sigma_8$ dataset shifts up the value of $w_a$ by almost $1\sigma$.

In the last part of Table \ref{tab:result}, we show results for the case of $w_0w_a$CDM background with a fixed $\gamma=0.55$. The constraints on $H_0$, $w_0$, $w_a$ and $S_8$ are nearly the same as those from the case with free $\gamma$. Therefore, although $f\sigma_8$ can improve the determination of $w_0$ and $w_a$ with respect to the PL18 dataset, as clearly seen in figure \ref{fig:trig-w0wa}, when adding background data from DESI and PP, its impact is quite small. 

A possible issue related to the use $f\sigma_8$ dataset, as used in ref. \citep{Nguyen:2023fip} and in our analysis, is the level of covariance these observations might have with distances measured with SNe Ia or galaxy surveys that derive BAO quantities, which in our work are included as PP or DESI datasets. The $f\sigma_8$ dataset has two types measurements: from peculiar velocities surveys, which use low redshift SNe Ia, e.g. \citep{Huterer:2016uyq,Boruah:2019icj}, and redshift space distortions, which use galaxies surveys data, e.g. \citep{Blake:2013nif,eBOSS:2020lta}. A detailed analysis of this issue is beyond the scope of this paper, but we proceed with the following consistency check. 

For the $w_0w_a$ background we also constrain the model with either DESI or PP datasets. As the results in table \ref{tab:result} indicate, constraints on $\gamma$ for PL18+$f\sigma_8$+DESI and PL18+$f\sigma_8$+PP are nearly the same. Moreover, when considering both DESI and PP, the improvement in the $\gamma$ determination with respect to datasets with only one of these background information is small. For the $\Lambda$CDM background, the differences in the determination of $\gamma$ is quite small when including DESI or DESI+PP over the PL18+$f\sigma_8$ dataset. These are valuable indications that the impact of the possible uncounted covariances should not be dramatic.

\begin{table*}
\centering
\footnotesize
\renewcommand{\arraystretch}{1.3}  
\setlength{\tabcolsep}{0.1pt}
\begin{tabular}{
    >{\raggedright\arraybackslash}m{2.7cm}
    >{\centering\arraybackslash}m{2cm}
    >{\centering\arraybackslash}m{2.5cm}
    >{\centering\arraybackslash}m{2.5cm}
    >{\centering\arraybackslash}m{2cm}
    >{\centering\arraybackslash}m{2cm}
    >{\centering\arraybackslash}m{2cm}
}
\hline 
Datasets & $H_0 [\text{km s}^{-1} \text{Mpc}^{-1}]$ & $\Omega_m$ & $S_{8}$ & $w_{0}$ & $w_{a}$ & $\gamma$\tabularnewline
\hline
\textbf{$\Lambda$CDM} &  &  &  &  &  & \tabularnewline
PL18 & $68.11\pm0.70$ & $0.3051\pm0.0092$ & $0.807\pm0.019$ & - - - & - - - & $0.664\pm0.068$\tabularnewline
PL18+DESI & $68.70\pm0.31$ & $0.2972\pm0.0039$ & $0.792\pm0.010$ & - - - & - - - & $0.700\pm0.057$\tabularnewline
PL18+$f\sigma_{8}$ & $67.91\pm0.49$ & $0.3076\pm0.0066$ & $0.813\pm0.011$ & - - - & - - - & $0.639\pm0.025$\tabularnewline
PL18+$f\sigma_{8}$+DESI & $68.53\pm0.29$ & $0.2994\pm0.0036$ & $0.8016\pm0.0078$ & - - - & - - - & $0.635\pm0.024$\tabularnewline
PL18+$f\sigma_{8}$ +DESI+PP & $68.44\pm0.28$ & $0.3006\pm0.0035$ & $0.8035\pm0.0075$ & - - - & - - - & $0.635\pm0.024$\tabularnewline
\hline
\textbf{$w_{0}w_{a}$CDM} &  &  &  &  &  & \tabularnewline
PL18 & $80.2^{+19}_{-7.3}$ & $0.240^{+0.022}_{-0.098}$ & $0.778^{+0.031}_{-0.048}$ & $-1.15^{+0.46}_{-0.68}$ & $<-0.249$ & $0.611^{+0.074}_{-0.093}$\tabularnewline
PL18+DESI & $64.6^{+1.9}_{-2.1}$ & $0.341\pm0.022$ & $0.828\pm0.019$ & $-0.54\pm0.22$ & $-1.31^{+0.73}_{-0.61}$ & $0.630\pm0.067$\tabularnewline
PL18+$f\sigma_{8}$ & $68.9\pm2.7$ & $0.300^{+0.021}_{-0.025}$ & $0.812\pm0.014$ & $-0.67^{+0.31}_{-0.21}$ & $<-1.01$ & $0.622^{+0.025}_{-0.029}$\tabularnewline
PL18+$f\sigma_{8}$+DESI & $65.9\pm1.6$ & $0.327\pm0.016$ & $0.822\pm0.012$ & $-0.67^{+0.15}_{-0.17}$ & $-0.98^{+0.51}_{-0.41}$ & $0.635\pm0.025$\tabularnewline
PL18+$f\sigma_{8}$+PP & $67.8\pm1.2$ & $0.308^{+0.011}_{-0.012}$ & $0.815\pm0.011$ & $-0.871\pm0.096$ & $-0.49\pm0.46$ & $0.634\pm0.026$\tabularnewline
PL18+$f\sigma_{8}$ +DESI+PP & $67.63\pm0.59$ & $0.3092\pm0.0056$ & $0.8128\pm0.0087$ & $-0.858\pm0.054$ & $-0.48^{+0.21}_{-0.18}$ & $0.633\pm0.024$\tabularnewline
CMB-P +$f\sigma_{8}$ & $61.6^{+2.8}_{-8.8}$ & $0.392^{+0.10}_{-0.064}$ & $1.02^{+0.22}_{-0.17}$ & $-0.53^{+0.40}_{-0.22}$ & $-1.0^{+1.1}_{-0.97}$ & $0.93^{+0.26}_{-0.32}$\tabularnewline
CMB-P +$f\sigma_{8}$+DESI & $63.4^{+1.5}_{-1.8}$ & $0.355\pm0.019$ & $0.934\pm0.061$ & $-0.40\pm0.19$ & $-1.77\pm0.57$ & $0.771\pm0.079$\tabularnewline
\hline
\textbf{$w_{0}w_{a}$CDM $(\gamma=0.55)$} &  &  &  &  &  & \tabularnewline
PL18 & $83.9^{+15}_{-5.6}$ & $0.214^{+0.015}_{-0.072}$ & $0.779^{+0.022}_{-0.046}$ & $-1.24^{+0.43}_{-0.57}$ & $<-0.561$ & - - -\tabularnewline
PL18+$f\sigma_{8}$ & $70.8\pm2.7$ & $0.282^{+0.020}_{-0.024}$ & $0.804\pm0.014$ & $-0.59^{+0.24}_{-0.14}$ & $<-1.79$ & - - -\tabularnewline
PL18+$f\sigma_{8}$+DESI & $66.1\pm1.6$ & $0.325^{+0.015}_{-0.017}$ & $0.823\pm0.012$ & $-0.68^{+0.15}_{-0.17}$ & $-0.99^{+0.52}_{-0.42}$ & - - -\tabularnewline
PL18+$f\sigma_{8}$ +DESI+PP & $67.72\pm0.59$ & $0.3083\pm0.0056$ & $0.8138\pm0.0088$ & $-0.855\pm0.053$ & $-0.50^{+0.21}_{-0.18}$ & - - -\tabularnewline
\hline
\end{tabular}
\caption{Marginalized 1D constraints and $68\%$ C.L. for the parameters of
each model.}
\label{tab:result}
\end{table*}

\subsection{Dependence of $\gamma$ on CMB data}
\begin{figure}
\centering
\includegraphics[width=0.6\linewidth]{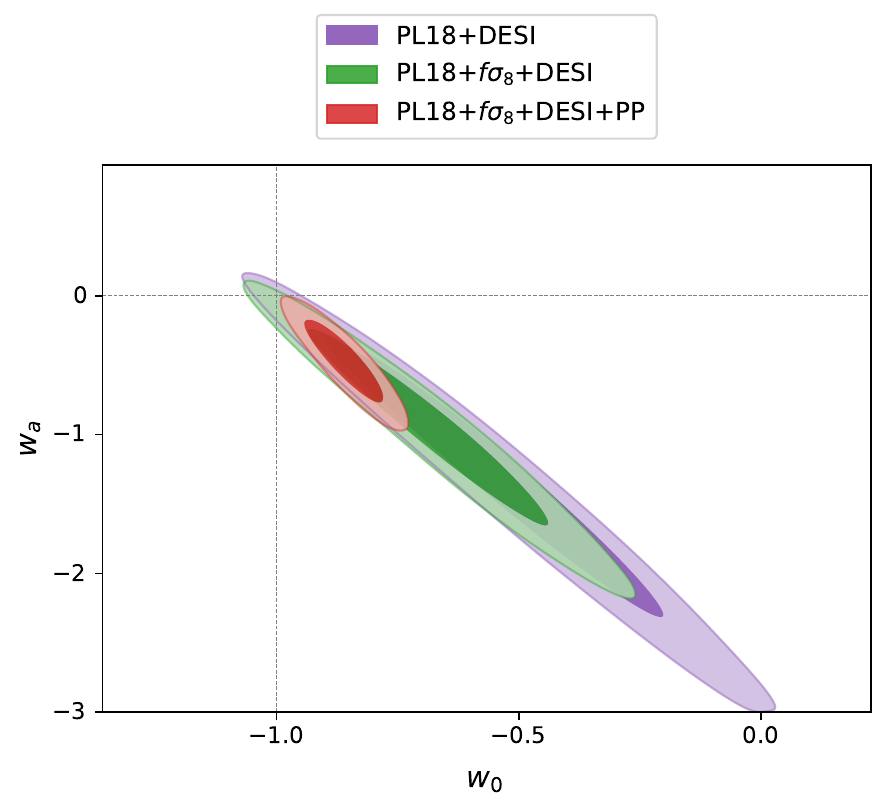}
\caption{Marginalized posteriors joint distributions for $w_{0}$ and $w_{a}$ parameters for selected datasets combinations. The contours enclose $68\%$ and $95\%$ of posteriors distributions. Dashed lines indicates $w_0=-1$ or $w_a=0$.}
\label{fig:w0wa-correlation}
\end{figure}

It was argued that, for the $\Lambda$CDM background, non-standard $\gamma$ values can be associated with lensing anomalies in CMB anisotropies \cite{Nguyen:2023fip,Specogna:2023nkq,Specogna2024, AtacamaCosmologyTelescope:2025nti}. Although this might be a concern for CMB studies, as our results demonstrate, PL18 data alone provide loose $\gamma$ constraints and the most relevant dataset for its determination is the $f\sigma_8$ one. Moreover, as can be seen in table \ref{tab:result} and in figure \ref{fig:gamma-dist}, for the $w_0w_a$CDM background, the inclusion of DESI and/or PP datasets prefers slightly higher values of $\gamma$ when compared to analysis with PL18 or PL18+$f\sigma_8$ datasets. Therefore, the determination of $\gamma$ is rather insensitive to the possible CMB anomalies associated with lensing effects. It is also worth of note that, as can be seen in figure \ref{fig:w0wa-correlation}, the inclusion of $f\sigma_8$ data improves the determination of $w_0$ and $w_a$, with respect to the case with PL18+DESI, essentially in the same region of the parameter space and correlation direction already allowed by this dataset, indicating a consistent determination also for these background parameters. 

\begin{figure*}
\centering
\includegraphics[width=\linewidth]{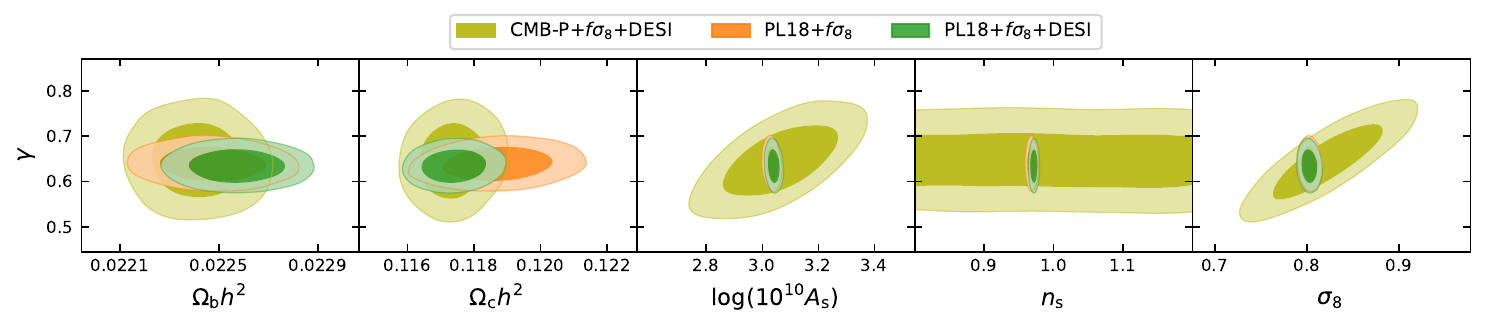}
\includegraphics[width=\linewidth]{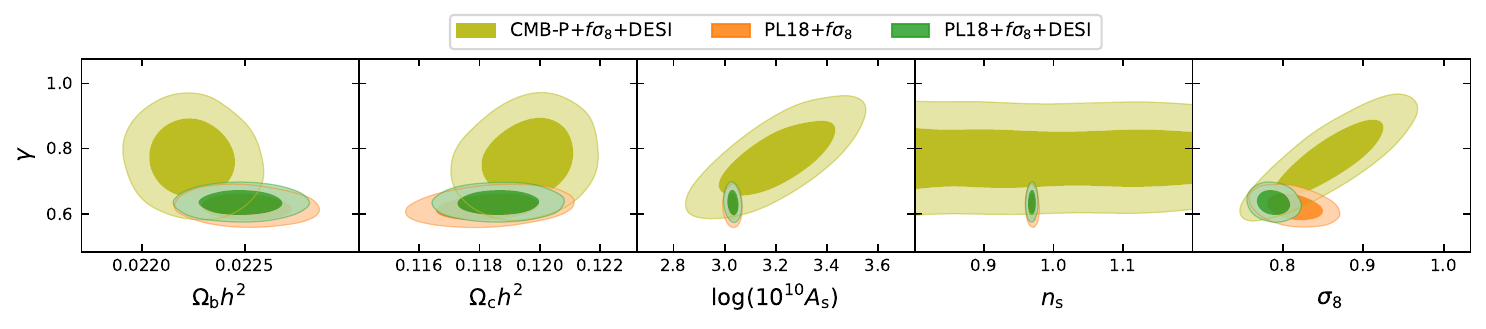}
\caption{Joint 1$\sigma $ and 2$\sigma $ posterior distributions for $\gamma$ and some selected parameters for $\Lambda$CDM (upper panel) and $w_{0}w_{a}$CDM (lower panel) backgrounds.}
\label{fig:CMB-P-constraints}
\end{figure*}

To better illustrate this point, we constrain $\gamma$ substituting the PL18 dataset by CMB priors \citep{Lemos:2023xhs}, which marginalizes out the dependence on lensing effects and integrated Sachs-Wolf effect including $f\sigma_8$ and DESI datasets. As can be seen in figure \ref{fig:CMB-P-constraints}, when using CMB priors, the determination of $A_s$ is degraded in comparison to the analysis that use PL18 dataset and $n_s$ is limited only by the prior. Even in this case, where the determination of $\sigma_8$ is also naturally degraded, we can see a clear preference for $\gamma>0.55$, excluding the standard value at $2.8\sigma$. Therefore, although CMB anisotropy spectra is important to constrain $A_s$, $n_s$, $\Omega_m$, which improves the precision of the $\gamma$ measurement around high values, the preference for non-standard growth index is fairly independent from late time effects on CMB that may be affected by astrophysical systematics or possible model dependent implementations for the lensing effects. In figure \ref{fig:CMB-P-constraints}, it is also interesting to note the positive correlation between $\gamma$ and $A_s$ or $\sigma_8$ that appears for the CMB-P+$f\sigma_8$+DESI analysis. This correlation indicates that higher $\gamma$, which suppresses the growth of matter perturbations, is partially compensated by higher $A_s$, with a consequential raising of $\sigma_8$. 

Also for these reasons, the different approaches to implement the $\gamma$ parametrizations discussed in \cite{Specogna:2023nkq} should have small effects in its determination. More specifically, the \texttt{CAMB\_ GammaPrime\_Growth} computes the impact of $\gamma$ on CMB anisotropies only in the lensing potential, while \texttt{MGCAMB} \citep{Wang:2023tjj}  also modifies the primary anisotropies, which are most sensitive to $\gamma $ at low-$\ell$, where cosmic variance is more relevant. 

Moreover, care must be taken when extrapolating the halo model \citep{Mead:2020vgs} to scenarios where it was not tested. In this context, analysis including high-$\ell$ data from ground based CMB experiments \citep{AtacamaCosmologyTelescope:2025nti,SPT-3G:2026uol}, which is more affected by lensing effects, is not consistent a priori. In particular, we remind that implementation of the $\gamma$ parametrization in the code \texttt{CAMB\_GammaPrime\_Growth} only considers its impact on the linear matter power spectrum that is used to compute the nonlinear matter power spectrum via the halo model, which in turn is used in the lensing kernel for CMB anisotropies. In other words, the halo model parameters that encode the physics of nonlinear evolution of perturbations is not yet fitted for models with growth history parametrized by $\gamma$.

\subsection{Model comparison}
In table \ref{tab:metrics} we present the $\Delta\chi^2$ of the $w_0w_a$CDM background model with free $\gamma$ with respect to case with fixed $\gamma=0.5519$, which is the theoretical value determined in \citep{Cortes2025} when constraining the background evolution with CMB priors and DESI datasets. As can be seen, allowing the growth index to vary improves all the fits to all considered datasets combinations. In this table, we also present the Bayes factor between the more complex model (with free $\gamma$), $M_1$, and the nested model (with $\gamma=0.5519$) $M_0$. Since the models are nested, we assume the Savage-Dickey density ratio,
\begin{equation}
    BF_{01} = \left.\frac{P(\gamma|d, M_1)}{P(\gamma|M_1)}\right|_{\gamma=0.5519},\label{eq:BF}
\end{equation}
where $d$ is a particular dataset, $P(\gamma|d, M_1)$ is the marginalized $\gamma$ distribution (as computed by \texttt{GetDist}) and $P(\gamma|M_1)=\mathcal{U}(0,2)$ is the prior distribution for $\gamma$, as given by table \ref{tab:priors}. 

Although the metric $\log_{10}BF_{10}$ indicates no preference for free $\gamma$ when using only the PL18 dataset, the inclusion of $f\sigma_8$ indicates a small preference. When including background data, the preference for the free growth index model becomes "strong" accourding to the Jeffrey's Scale \citep{10.1093/oso/9780198503682.001.0001}, i.e., $\log_{10} BF_{10}>1$. In last column of table \ref{tab:metrics} we also present the significance of the tension between the obtained values and theoretical prediction in unis of standard deviations for Gaussian distributions. As can be noted, the inclusion of background data increases the $\gamma$ tension to about $3.3\sigma$. 

As discussed earlier, the theoretical $\gamma$ values have some dispersion. We checked that setting $\gamma=0.5519+0.0011$ for the model with fixed $\gamma$, where the added value is the theoretical $1\sigma$ value, does not reduce the evidence in favor of free $\gamma$ model substantially, still keeping $\log_{10} BF_{10}>1$ for the two more restrictive datasets. 

\begin{table}
\centering
\begin{tabular}{lccc}
\hline
Datasets & $\Delta\chi^{2}$ & $\log_{10} BF_{10}$ & Tension [$\sigma$] \tabularnewline
\hline
PL18                       & $-4.9$  & $-0.91$ & $0.63$\tabularnewline
PL18+$f\sigma_{8}$         & $-7.4$  & $+0.21$ & $2.41$\tabularnewline
PL18+$f\sigma_{8}$+DESI    & $-12.2$ & $+1.24$ & $3.32$\tabularnewline
PL18+$f\sigma_{8}$+DESI+PP & $-10.6$ & $+1.15$ & $3.37$\tabularnewline
\hline
\end{tabular}
\caption{$\Delta\chi^{2}\equiv\chi^{2}_{\gamma}-\chi^{2}_{\gamma=0.5519}$, $\log_{10} BF_{10}$ and the tension as $(\gamma-0.5519)/\sigma_\gamma$ for the $w_{0}w_{a}$CDM background and some selected datasets.}
\label{tab:metrics}
\end{table}

We also briefly analyze the case of clustering DE (CDE). As shown in ref. \citep{Cortes2025}, when constraining the $w_0w_a$CDM background with CMB priors and DESI data and assuming null sound speed for DE perturbations, the theoretical prediction is 
\begin{equation}
    \gamma _{w_0w_a,CDE}= 0.5644 ^{+0.0043} _{-0.0028}.
    \label{eq:gamma-CDE}
\end{equation}
As explained in ref. \citep{Cortes2025}, this slightly higher value is a consequence of the transient phantom behavior preferred by DESI data. In this scenario, the tension between the measured value and the theoretical value decreases to $2.8\sigma$. 

\section{Conclusions\label{sec:conclusion}}
Our main conclusion is that the determination of $\gamma$ for the $w_0w_a$CDM background is quite similar to the values already obtained for the $\Lambda$CDM one. The best determinations in our analysis, obtained with the PL18+$f\sigma_8$+DESI+PP dataset, are:
\begin{equation}
  \gamma_{\Lambda}=0.635\pm0.024 \mbox{ and }  \gamma_{w_0w_a}=0.633\pm0.024\,.
\end{equation}
In our study, we found $3.5\sigma$ tension for $\Lambda$CDM and $3.4\sigma$ $w_0w_a$CDM background. Therefore, even in a background model with two extra parameters for the DE EoS with respect to the case for $\Lambda$CDM background, the $\gamma$ tension is barely alleviated. Previous analysis, with other dataset combinations, found $3.7\sigma$ and $3.5\sigma$ for $\Lambda$CDM \citep{Nguyen:2023fip,AtacamaCosmologyTelescope:2025nti} respectively. 

We also found that the determination of $\gamma$ is consistent across many datasets combinations and that the most relevant information comes from the $f\sigma_8$ measurements. In particular, we showed that the PL18 dataset provides loose constraints on $\gamma$ and, by substituting this dataset by CMB priors, high values of $\gamma$ are still preferred, indicating that possible systematics issues with CMB lensing are not responsible for the $\gamma$ tension. Therefore, if one seeks for unaccounted systematics that might by yielding $\gamma$ from the expected values in $\Lambda$CDM or $w_0w_a$CDM backgrounds, the right place is on the $f\sigma_8$ dataset.

Future data can be decisive in confirming or excluding the $\gamma$ tension. The Euclid mission expects to measure it with uncertainty bellow $0.02$ \citep{EuclidCollaboration2024}. If a lower $\gamma$ is found, but still with some tension with the $\Lambda$CDM value,  it can be interpreted as an indication of clustering DE, which can provide $\gamma\simeq0.56$  in the case its EoS is phantom for a relevant period \citep{Cortes2025}. In the case $\gamma\simeq0.63$ is confirmed, modified gravity models should be considered as an theoretical explanation.
In particular, phenomenological parametrizations for a varying effective Newton's constant can readily accommodate $\gamma\simeq 0.63$ \citep{Cortes:2025wni}, implying a weakening of the effective gravitational interaction at late times. However, many modified gravity theories, such as $f(R)$ gravity and the normal branch of DGP, typically predict an enhanced gravitational coupling, leading to faster structure growth. A weaker effective coupling can instead arise in specific scenarios, such as the self-accelerating branch of DGP and certain quartic Galileon models \cite{Li2013,Li2013a}, although these models face significant theoretical or observational constraints. More recently, a scalar-field model with momentum exchange between dark energy and dark matter has been proposed, allowing for a reduced effective gravitational coupling in the cold dark matter sector and suppressed structure growth \citep{Pookkillath:2026wyg}.

\acknowledgments

We thank Valerio Marra, Miguel Quartin and David Camarena for usefull discutions. RCB and IBSC thanks Conselho Nacional de Desenvolvimento Científico e Tecnológico–CNPq (process 444368/2024-8) for partially supporting this work. IBSC thanks Universidade Federal do Rio Grande do Norte for the undergraduate research scholarship No. 04/2026 (PIBIC-PROPESQ) project  PIJ25145-2026. LGM also thanks Conselho Nacional de Desenvolvimento Científico e Tecnológico-CNPq (grant 305831/2026-7) for partial financial support. This work was developed using the computational resources of Laboratório de Multiprocessamento Computacional e Científico, from Escola de Ciências e Tecnologia of Universidade Federal do Rio Grande do Norte.

\bibliographystyle{JHEP}
\bibliography{referencias.bib}
\end{document}